\documentclass[a4paper,twocolumn,fleqn]{article}

\usepackage{ist}
\usepackage{siunitx} % package for SI units
\usepackage{float} % package for float specifier "H"
\usepackage{algorithm, algpseudocode} % algorithms listing
\usepackage[capitalise]{cleveref} %Referencing without need to explicitly state fig /table

\title{Digital Pre-Distorted One-Step Phase Retrieval Algorithm for Real-Time Hologram Generation for Holographic Displays}

\author{Jinze Sha, Adam Goldney, Andrew Kadis, Jana Skirnewskaja, Timothy D. Wilkinson}

\date{} % date has an empty field.

\begin{document}

\maketitle

\thispagestyle{empty} % prevents the first page to be numbered

%%%%%%%%%%%%%%%%%%%%%%%%%%%%%%%%%%
% Abstract
%%%%%%%%%%%%%%%%%%%%%%%%%%%%%%%%%%

\begin{abstract}
    In a computer-generated holographic projection system, the image is reconstructed via the diffraction of light from a spatial light modulator. In this process, several factors could contribute to non-linearities between the reconstruction and the target image. This paper evaluates the non-linearity of the overall holographic projection system experimentally, using binary phase holograms computed using the one-step phase retrieval (OSPR) algorithm, and then applies a digital pre-distortion (DPD) method to correct for the non-linearity. Both a notable increase in reconstruction quality and a significant reduction in mean squared error were observed, proving the effectiveness of the proposed DPD-OSPR algorithm.
\end{abstract}

%%%%%%%%%%%%%%%%%%%%%%%%%%%%%%%%%%%%
% Body
%%%%%%%%%%%%%%%%%%%%%%%%%%%%%%%%%%%%
\section{Introduction}
In a computer-generated holographic projection system, images are generated via the controlled diffraction of coherent light, which is modulated by a spatial light modulator (SLM). Contemporary SLM's can only modulate either phase or amplitude, hence algorithms are needed to compute amplitude-only or phase-only holograms. Among those, phase-only holograms are generally preferred due to inherently higher energy efficiency as there is no intentional blockage of light during modulation. Classic phase retrieval algorithms include direct binary search \cite{Seldowitz1987}, simulated annealing \cite{Kirkpatrick1983}, Gerchberg-Saxton \cite{Gerchberg1972}, and recent years have also seen numerical optimization methods \cite{Zhang2017, Liu2020, Choi2021, Chen2021, Kadis2022}, but these methods are iterative so computation speed is the major challenge. The previous research had demonstrated a real-time computer-generated holography (CGH) method called one-step phase retrieval (OSPR) \cite{Cable2004}, which was fast enough for real-time holography, but the reconstruction quality still has potential for improvement. Hence, a computationally inexpensive method is needed to improve the reconstruction quality whilst maintaining the real-time property of the OSPR algorithm.

There are several factors contributing to the non-linearities between the reconstruction of hologram and the target image, including the calculation and quantization of the hologram, the modulation of the light and the imperfections in the optical setup. This article proposes the digital pre-distorted one-step phase retrieval (DPD-OSPR) algorithm. The digital pre-distortion (DPD) is carried out on the holographic projection system using holograms computed by the OSPR algorithm by measuring the non-linearity experimentally and applying the according pre-distortion curve on target images. DPD can be done via a one-to-one correction curve or a look-up table (LUT) which allows the relationship between the input and output to be adjusted without any heavy computation.

The intuition of the proposed DPD-OPSR algorithm for CGH comes from the gamma correction method for conventional displays, such as cathode-ray tube (CRT) monitor \cite{Xu:09}, plasma display panel television (PDP-TV) \cite{Sung:09} and thin-film-transistor liquid-crystal display (TFT LCD) \cite{Lee:05,Prraga:14}. Gamma correction for conventional displays were originally developed to mimic the perceptual response of human vision \cite{Poynton2012}. The work presented here is a logical continuation of this approach applied to holographic displays.

\section{Method}
The DPD-OSPR method builds on the OSPR algorithm \cite{Cable2004}. The OSPR algorithm relies on the time multiplexing of holograms, exploiting the response time of eyes in order to reduce noise in the replay field, rather than computational optimization of the hologram. As the random noises are averaged by the eye while the target image stays, the perceived noise is lessened by the temporal average detected by the eye \cite{Cable2006}.
\begin{algorithm}[H]
    \caption{One-Step Phase Retrieval (OSPR) algorithm}\label{alg:One Step Phase Retrieval (OSPR) Algorithm}
    \textbf{Input:} Target field $T$, Propagation function $\mathcal{P}$, Number of sub-frames $S$ \\
    \textbf{Output:} List of phase holograms $H[1\ldots S]$
    \begin{algorithmic}
        \State // Compute a list of hologram sub-frames based on different additive random phase
        \For {$s$ = $1$ to $S$}
        \State $E \gets T * $ RandomPhase()
        \State $A \gets \mathcal{P}^{-1}[E]$
        \State $H[s] \gets \angle A$
        \EndFor\\
        \State // Then display the sub-frames on the SLM sequentially
        \State $s\gets 1$
        \While {True}
        \State Display($H[s]$)
        \State $s\gets s + 1$
        \If {$s > S$}
        \State $s\gets 1$
        \EndIf
        \EndWhile
    \end{algorithmic}
\end{algorithm}
The OSPR algorithm is described in \cref{alg:One Step Phase Retrieval (OSPR) Algorithm}, where the propagation function $\mathcal{P}$ is simply the Discrete Fourier Transform (DFT) for Fraunhofer diffraction \cite{Daintith2009}. The number of sub-frames $S$ is set to 24 for this article. The OSPR algorithm generates each hologram sub-frame by taking the inverse Discrete Fourier Transform (iDFT) on the field concatenating the target image with a random phase, and then the phase-only constraint is applied by discarding the amplitude while keeping the phase. After computing the $S$ subframes, they are then displayed on the SLM sequentially. For real-world SLM with limited bit depth, a further step of quantization on each subframe is needed.

\section{Experimental setup}

\begin{figure}[H]
    \centering
    \includegraphics[width=0.45\textwidth]{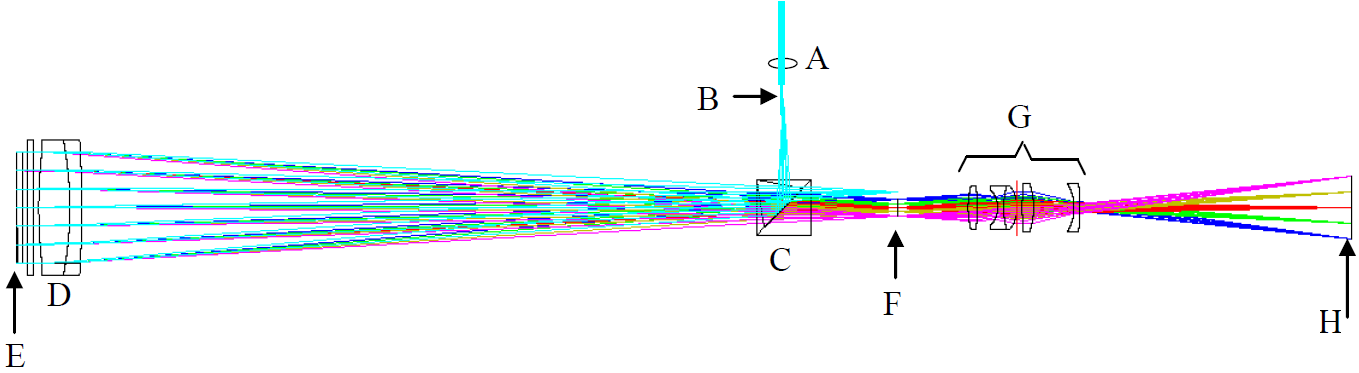}
    \caption{Optical setup \cite{Freeman2009}}
    \label{fig:projection_setup}
\end{figure}

The holographic projector used in this experiment is a Fourier projection system developed by Freeman \cite{Freeman2009}, as shown in \cref{fig:projection_setup}. The design is consisted of a diode-pumped solid-state (DPSS) 532 nm 50mW laser source, focussed down by an aspheric singlet (A), the focus of which becomes the diffraction limited point source (B) for the projector. The beam then passes through a polarizing beam splitter cube (C) to a collimating lens (D), which illuminates the SLM (E). The SLM is a binary phase SXGA-R2 ForthDD ferroelectric Liquid crystal on silicon (LCOS) micro-display with a refresh rate of 1440Hz, a pixel pitch of \SI{13.6}{\micro\metre} and a resolution of $1280\times1024$. An aperture at point (F) spatially filters out the other orders, leaving only one first order, which is then magnified up by a finite conjugate lens group (G) to produce an image, of the required size, on the screen (H). \cite{Freeman2009}

\begin{figure}[H]
    \centering
    \includegraphics[width=0.45\textwidth]{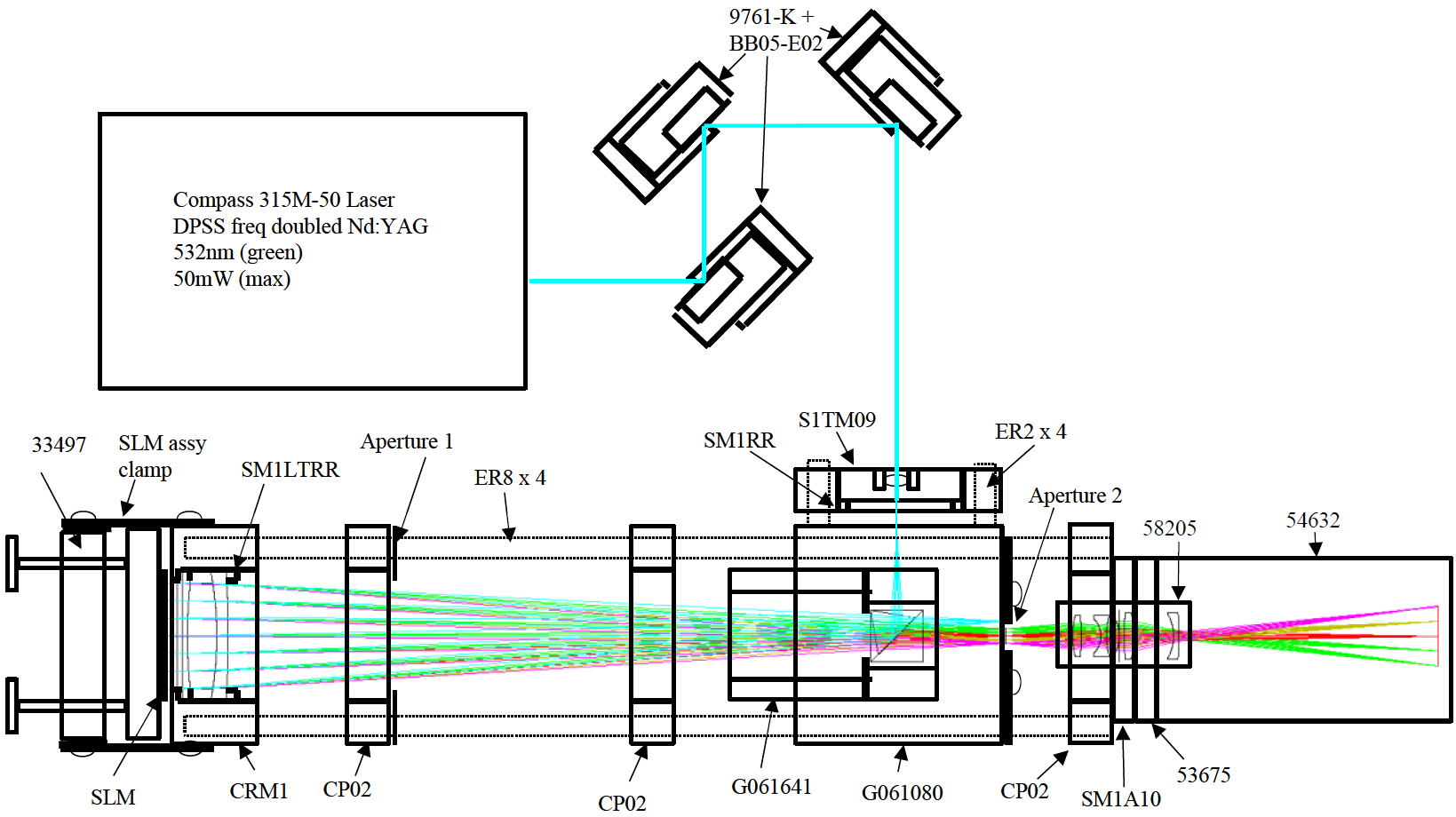}
    \caption{Mechanical components \cite{Freeman2009}}
    \label{fig:holographic_projector}
\end{figure}

The mechanical components are listed in \cref{fig:holographic_projector} with parts number. The holograms displayed on the SLM are generated using the OSPR algorithm \cite{Cable2004}, and as the SLM is a binary-phase modulator, each hologram sub-frame needs to be binary quantized. Then each group of the 24 binary-phase hologram sub-frames are encoded as the 8-bit reg, green, blue (RGB) channels of a 24-bit image to interface with the SLM driver electronics. The SLM displays each bit plane sequentially, with ones and zeros mapping to opposing phase modulations at each pixel. The reconstructions were captured using a Canon 550D camera with an EFS 18-55 mm lens. To ensure fair comparison, the camera was set to the same manual setting when comparing each pair of replay fields before and after DPD. It takes $24/1440=1/60$s to display all 24 sub-frames on a 1440Hz SLM, so the camera shutter speed was set to $1/30$s to capture all frames twice. The images captured are in 24-bit RGB colour, which are converted to grey-scale in 8-bit depth when calculating normalized mean squared error (NMSE).

\section{Determining the DPD curve}

\begin{figure}[H]
	\centering
	\includegraphics[width=0.45\textwidth]{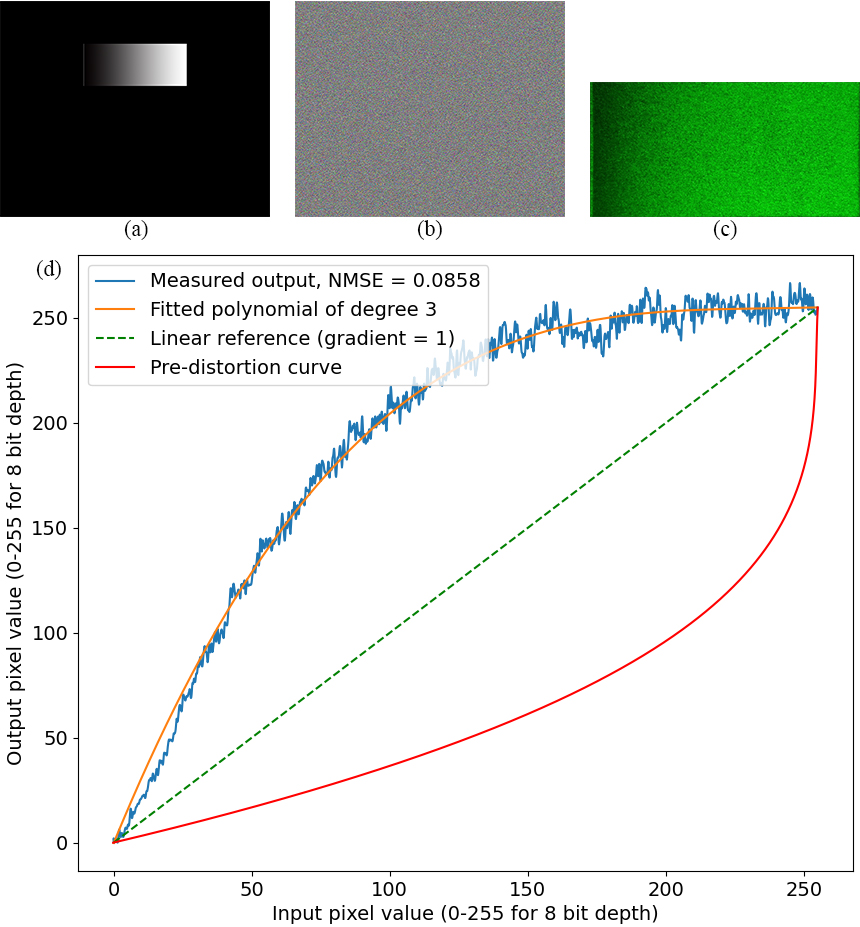}
	\caption{Determining the DPD curve. (a) Input linear grey-scale ramp. (b) Corresponding CGH of (a) with 24-subframe binary phase encoding. (c) Holographic projection replay field of (b). (d) Plot of non-linearity measurement and according pre-distortion curve.}
	\label{fig:Determining_the_DPD_curve}
\end{figure}

To determine the DPD curve of the holographic projection system, the non-linearity needs to be measured first. The hologram in \cref{fig:Determining_the_DPD_curve}(b) was first generated using OSPR algorithm for the linear grey-scale ramp of pixel value increasing linearly from 0 to 255, as shown in \cref{fig:Determining_the_DPD_curve}(a), along with a single pixel white (255) strip at the left end as a fiducial marker to demonstrate the beginning of the grey-scale region \cite{Cable2006}.

The projection output of the linear grey-scale ramp was then captured and cropped as shown in \cref{fig:Determining_the_DPD_curve}(c), from which the non-linearity curve was determined, by averaging each column of pixels in the image and discarding the fiducial marker, forming the blue line in \cref{fig:Determining_the_DPD_curve}(d). A third-order polynomial fit was applied, generating a smoothed non-linearity curve (yellow line in \cref{fig:Determining_the_DPD_curve}(d)).

There exhibits a high degree of non-linearity. By taking the mean of the square of the error between the measured output (blue line) and the linear reference (green dashed line), the normalized mean squared error (NMSE) of the measured output was calculated to be 0.0858. To correct for the non-linearity, the DPD curve (red line) was formed by inverting the smoothed non-linearity (yellow line) in \cref{fig:Determining_the_DPD_curve}(d).

\begin{figure}[H]
	\centering
	\includegraphics[width=0.45\textwidth]{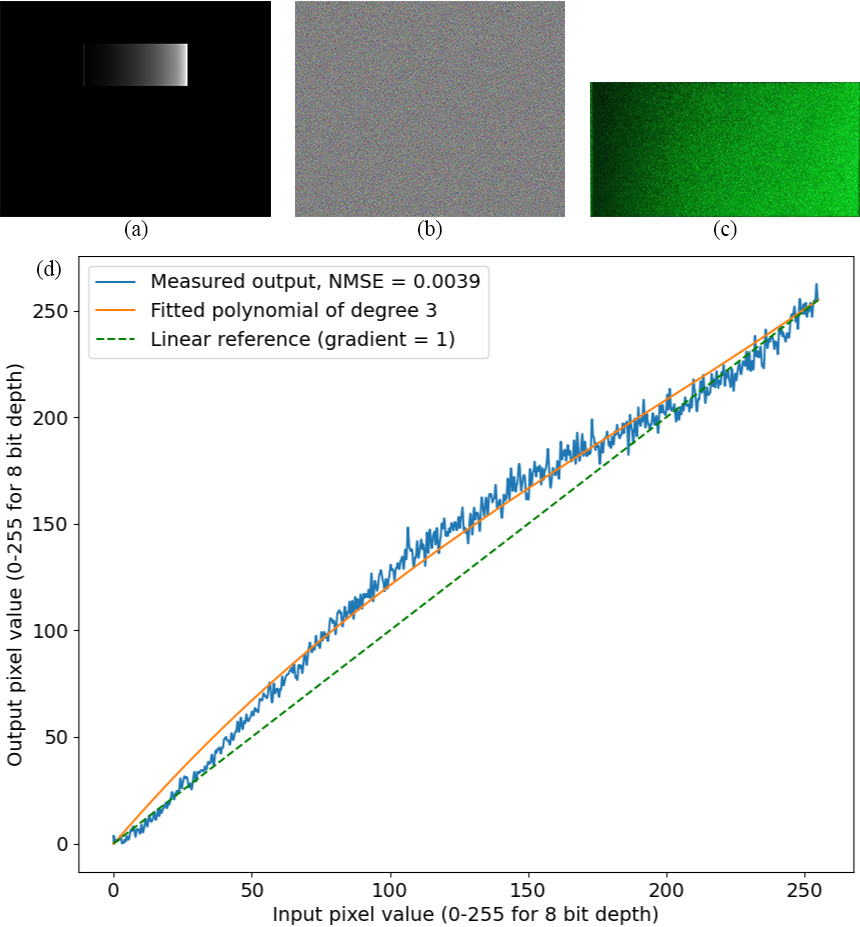}
	\caption{Validation of DPD curve on the grey-scale ramp. (a) Pre-distorted ramp. (b) Corresponding CGH of (a) with 24-subframe binary phase encoding. (c) Holographic projection replay field of (b). (d) Non-linearity measurement after DPD.}
	\label{fig:Validation of DPD curve on the grey-scale ramp}
\end{figure}
\vspace{3mm}

Subsequently, the DPD curve (red line in \cref{fig:Determining_the_DPD_curve}(d)) was used to adjust the grey-scale ramp, achieving the pre-distorted grey-scale ramp as shown in \cref{fig:Validation of DPD curve on the grey-scale ramp}(a). The according projection output was then captured as shown in \cref{fig:Validation of DPD curve on the grey-scale ramp}(c). By using the same method of averaging columns of pixels, the measured output was plotted in \cref{fig:Validation of DPD curve on the grey-scale ramp}(d). It can be seen that the corrected non-linearity was much closer to linear comparing to the original non-linearity, and the NMSE was calculated to be 0.0039.

\begin{table}[H]
    \begin{center}
        \begin{tabular}{|c|c|c|}
            \hline
                       & NMSE   & Percentage \\ \hline
            Before DPD & 0.0858 & 100\%      \\ \hline
            After DPD  & 0.0039 & 4.55\%     \\ \hline
        \end{tabular}
        \caption{Non-linearity results before and after DPD}
        \label{tab:non-linearity result}
    \end{center}
\end{table}

Hence, as demonstrated in \cref{tab:non-linearity result}, DPD achieved a 95.45\% reduction in MSE, which was a significant improvement in non-linearity, therefore the DPD curve measured is validated.

\section{Applying the DPD Curve}

\begin{figure}[H]
	\centering
	\includegraphics[width=0.45\textwidth]{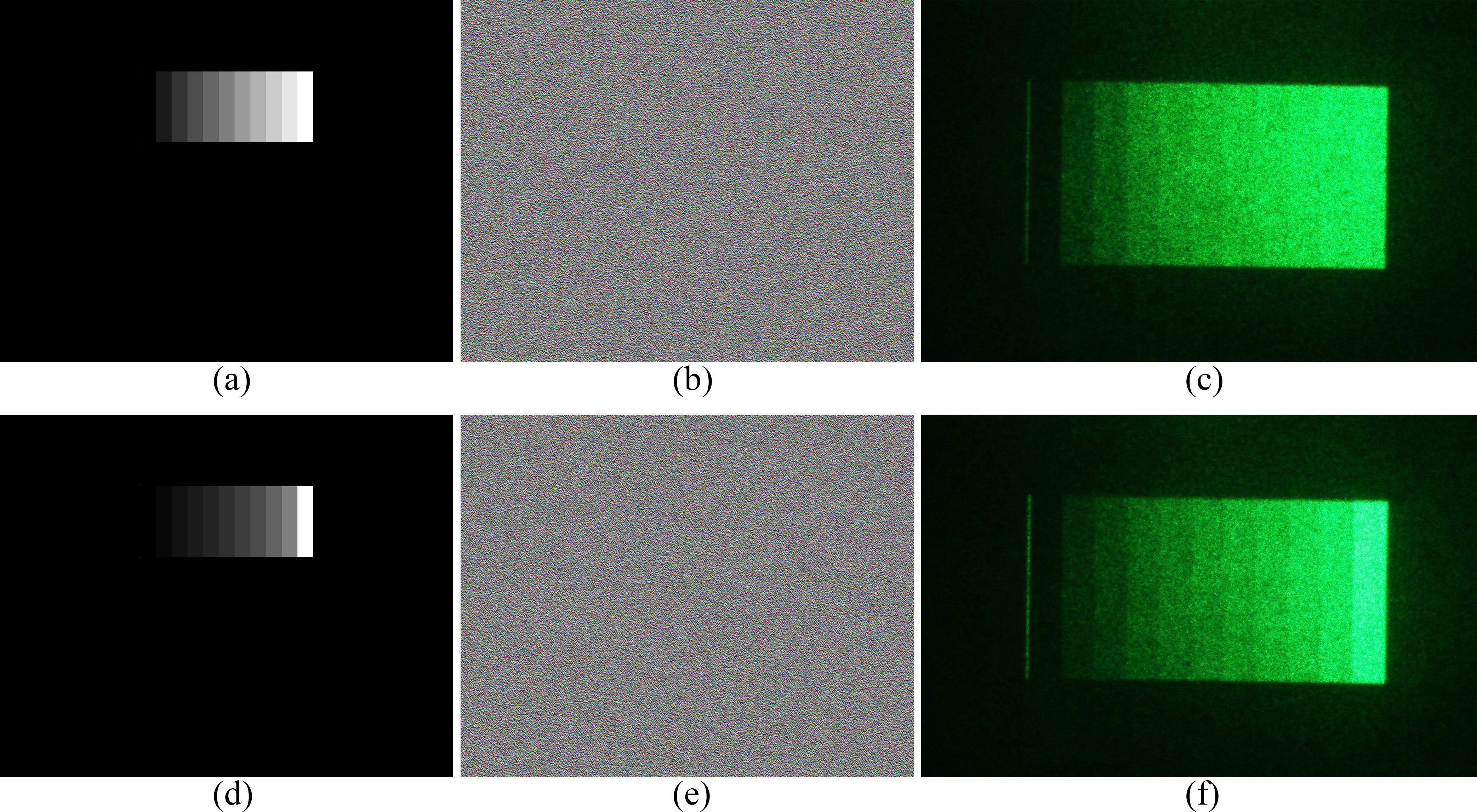}
	\caption{Application of DPD on the 10-step strips. (a) 10 strips with equal step of pixel value. (b) CGH of (a). (c) Holographic projection replay field of (b). (d) After DPD of (a). (e) CGH of (d). (f) Holographic projection replay field of (e).}
	\label{fig:10_step_strips}
\end{figure}
\vspace{3mm}

To qualitatively demonstrate the effectiveness of our approach, we project a simple test pattern of a graduated ramp test pattern consisted of 10-step strips in \cref{fig:10_step_strips}(a), which is commonly employed in gamma-correction calibration of many display systems. As shown in the projection replay field captured in \cref{fig:10_step_strips}(c), before DPD, the right few strips are barely distinguishable. In comparison, after carrying out DPD, it can be seen that each pair of adjacent strips in \cref{fig:10_step_strips}(f) are much more distinguishable, qualitatively showing the effectiveness of the DPD method.

\begin{figure}[H]
	\centering
	\includegraphics[width=0.45\textwidth]{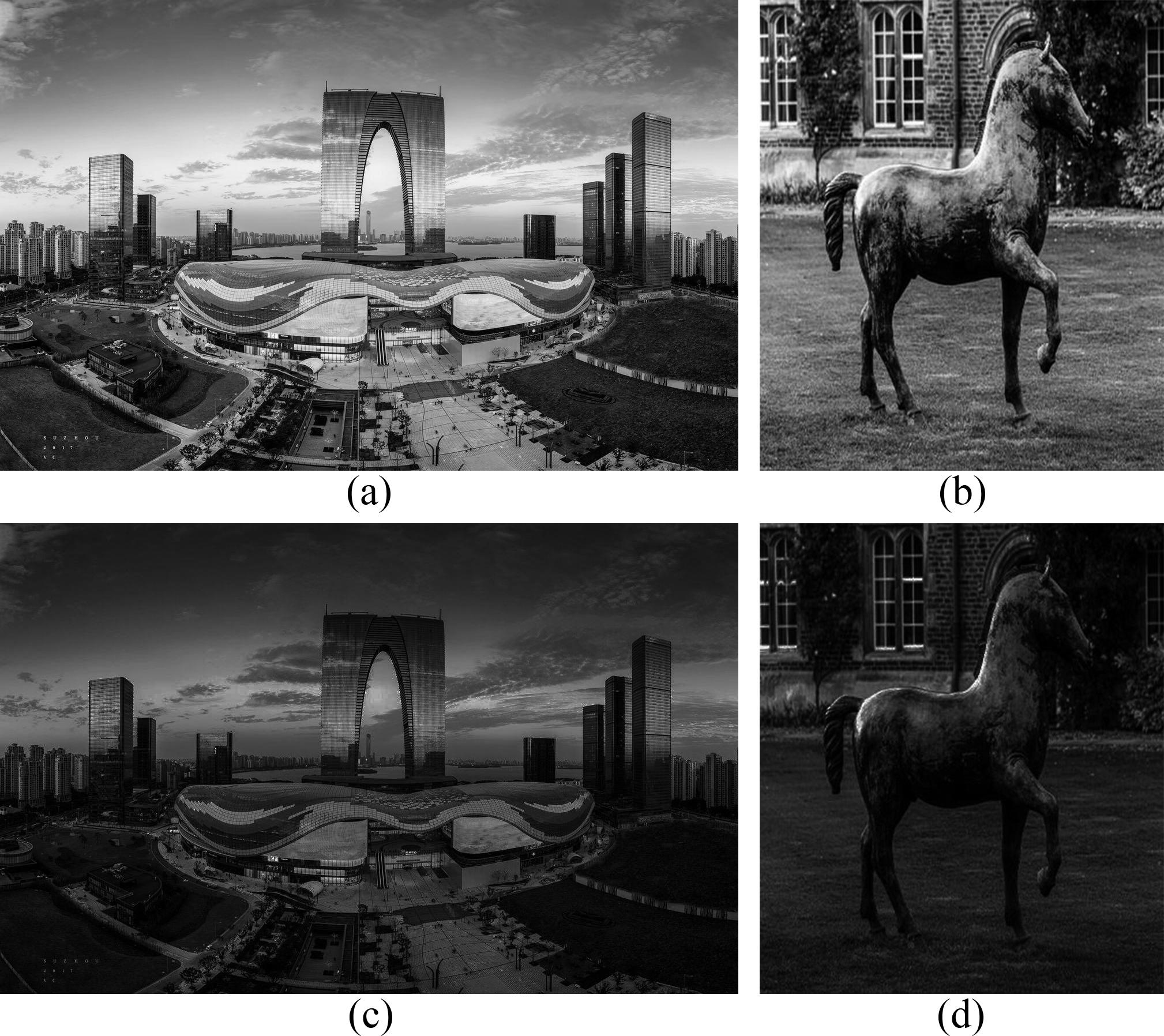}
	\caption{Application of DPD on two sample real-word image. (a) Sample image 1: City Scene \cite{Zhao2017}. (b) Sample image 2: Horse. (c) Sample image 1 after DPD. (d) Sample image 2 after DPD.}
	\label{fig:Application of DPD on two sample real-word image}
\end{figure}
\vspace{3mm}

\begin{figure}[H]
	\centering
	\includegraphics[width=0.45\textwidth]{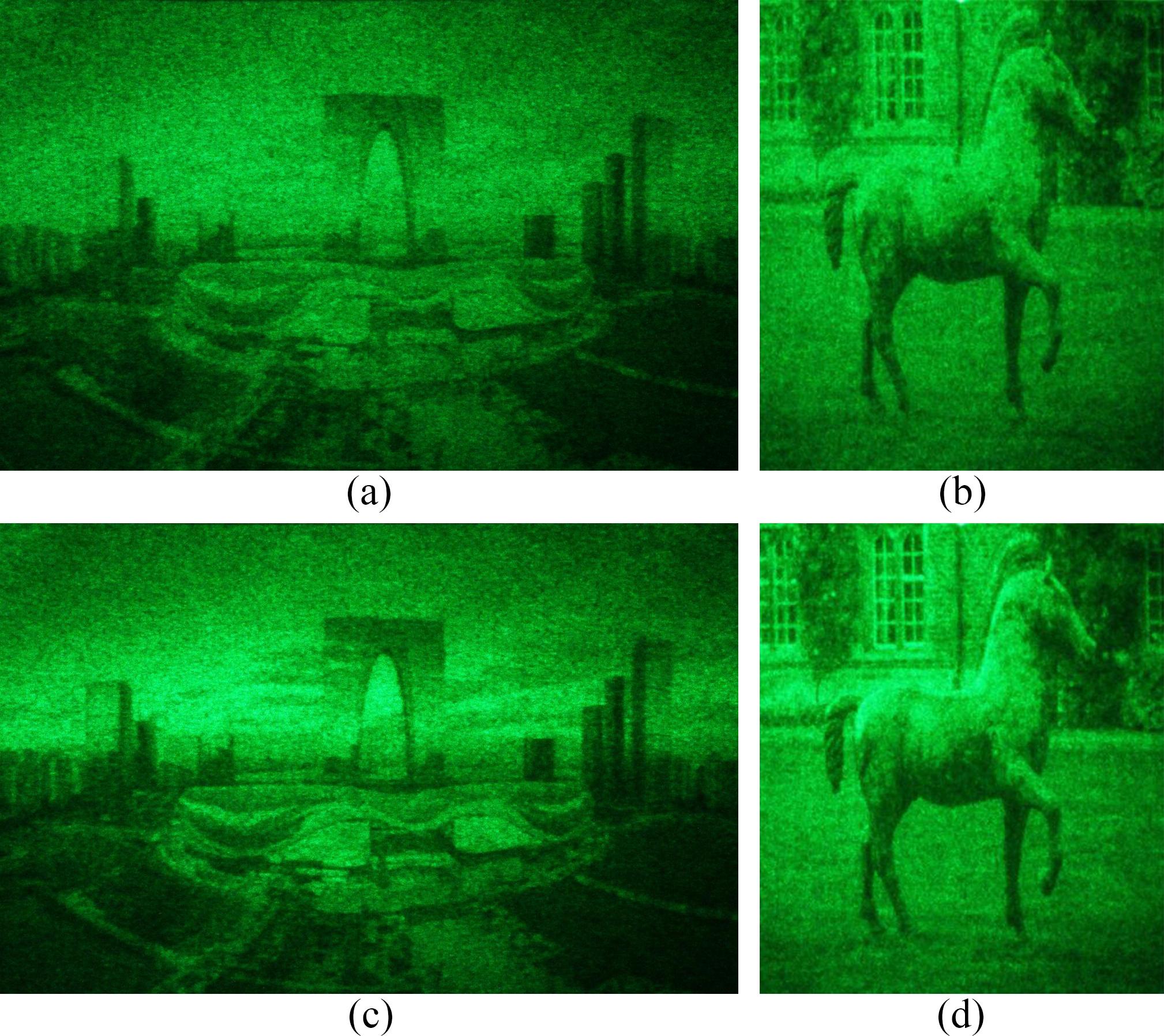}
	\caption{Projection output of the two sample images before and after DPD. (a) Replay field of Sample image 1 before DPD (NMSE=0.06139). (b) Replay field of Sample image 2 before DPD (NMSE=0.04309). (c) Replay field of Sample image 1 after DPD (NMSE=0.04920). (d) Replay field of Sample image 2 after DPD (NMSE=0.03635).}
	\label{fig:Projection output of the two sample images before and after DPD}
\end{figure}

Then the DPD curve was applied to the two sample images as shown in \cref{fig:Application of DPD on two sample real-word image} (a) and (b), producing pre-distorted images in \cref{fig:Application of DPD on two sample real-word image} (c) and (d). Holograms were generated for each image using the OSPR algorithm and loaded onto the SLM respectively. The according replay fields were captured as shown in \cref{fig:Projection output of the two sample images before and after DPD}.

\vspace{3mm}

The replay fields of the holographic projection of original images are shown in \cref{fig:Projection output of the two sample images before and after DPD} (a) and (b), and the replay fields of the holographic projection of images after DPD are shown in \cref{fig:Projection output of the two sample images before and after DPD} (c) and (d).

As shown in \cref{fig:Projection output of the two sample images before and after DPD}(a), it can be seen that, before DPD, the edges between the buildings and the sky were quite ambiguous, with most detail of the sky being lost. In comparison, after DPD, the replay field in \cref{fig:Projection output of the two sample images before and after DPD}(c) provided not only sharper edges between buildings and the sky, but also more detail of clouds in the sky. The NMSE of the replay field for sample image 1 decreased from 0.06139 to 0.04920, which was a 19.86\% reduction.

In \cref{fig:Projection output of the two sample images before and after DPD}(b), before DPD, the horse was difficult to distinguish from the background, especially around the horse's back area. But after DPD, as shown in \cref{fig:Projection output of the two sample images before and after DPD}(d), contrast has been significantly boosted and the fine detail around this part of the horse is more evident. The NMSE of the replay field for sample image 2 decreased from 0.04309 to 0.03635, which was a 15.64\% reduction.

\begin{table}[H]
    \begin{center}
        \begin{tabular}{ccc}
            \hline
            \multicolumn{1}{|c|}{Sample image 1} & \multicolumn{1}{c|}{NMSE}    & \multicolumn{1}{c|}{Percentage} \\ \hline
            \multicolumn{1}{|c|}{Before DPD}     & \multicolumn{1}{c|}{0.06139} & \multicolumn{1}{c|}{100\%}      \\ \hline
            \multicolumn{1}{|c|}{After DPD}      & \multicolumn{1}{c|}{0.04920} & \multicolumn{1}{c|}{80.15\%}    \\ \hline\hline
            \multicolumn{1}{|c|}{Sample image 2} & \multicolumn{1}{c|}{MSE}     & \multicolumn{1}{c|}{Percentage} \\ \hline
            \multicolumn{1}{|c|}{Before DPD}     & \multicolumn{1}{c|}{0.04309} & \multicolumn{1}{c|}{100\%}      \\ \hline
            \multicolumn{1}{|c|}{After DPD}      & \multicolumn{1}{c|}{0.03635} & \multicolumn{1}{c|}{84.36\%}    \\ \hline
        \end{tabular}
        \caption{DPD results for sample images}
        \label{tab:DPD results for sample images}
    \end{center}
\end{table}

Hence, as summarised in \cref{tab:DPD results for sample images}, DPD achieved a 19.86\% reduction in NMSE for sample image 1 and a 15.64\% reduction in NMSE for sample image 2, quantitatively proving the effectiveness of DPD method for CGH of real-world test images using OSPR algorithm.

Lastly, as the DPD is a one-to-one mapping, the computation time is negligible. In practice, the computational overhead is too small to be measured against randomness between subsequent runs. DPD can also be further accelerated in hardware using a hardware LUT, so that the DPD can be carried out instantly. This approach is widely adopted in gamma correction for displays.

\section{Conclusion}
The non-linearity between target image and reconstructed image was measured for the overall holographic projection system by projecting a linear grey-scale ramp. Then DPD was applied to the grey-scale ramp and successfully reduced the MSE by 95.45\%. To examine its effectiveness on real world images, the DPD method was applied on two sample images, it was observed that more details were shown in the replay field after DPD, and the MSE's of the two example images were reduced by 19.86\% and 15.64\%. As the DPD is a one-to-one mapping, the extra computation required is negligible. Hence, we have demonstrated the effectiveness of the proposed DPD-OSPR method to improve reconstruction quality on the existing OSPR algorithm while still keeping its ability for real-time holography.

\section{Acknowledgments}
This work was supported by the Engineering and Physical Sciences Research Council (EPSRC) [EP/S022139/1].


%%%%%%%%%%%%%%%%%%%%%%%%%%%%%%%%%%
% Bibliography
%%%%%%%%%%%%%%%%%%%%%%%%%%%%%%%%%%

\small
\bibliographystyle{ieeetr}
\bibliography{gc-references}


\begin{biography}
    Jinze Sha received his BA and MEng from the University of Cambridge (June 2020). After graduation, he worked for Advanced Micro Devices (AMD) for a year (Aug 2020- Aug 2021). Then he returned to the University of Cambridge to pursue a PhD degree (from Oct 2021). His research has focused on algorithms for computer generated holography (CGH).
\end{biography}

\end{document}